\documentclass[reqno]{article}

\usepackage{amsmath,amsthm,amssymb}

\newtheorem{theorem}{Theorem}
\newtheorem{lemma}{Lemma}
\newtheorem{corollary}[theorem]{Corollary}
\newtheorem{problem}{Problem}
\newtheorem{claim}{Claim}[theorem]

\def\crit{\mbox{Crit}}

\begin{document}
\title{Maximum directed cuts in digraphs with degree restriction}

\author{ Jen\H o Lehel \thanks{%
Department of Mathematical Sciences, The University of Memphis,
Tennessee, USA and Computer and Automation Research Institute of the
Hungarian Academy of Sciences, Budapest, Hungary; jlehel@memphis.edu}
\and%
Fr\'ed\'eric Maffray \thanks{%
CNRS, Laboratoire G-SCOP, 46 avenue F\'elix Viallet, 38031 Grenoble
Cedex, France; frederic.maffray@g-scop.inpg.fr}
\and%
Myriam Preissmann \thanks{%
CNRS, Laboratoire G-SCOP, 46 avenue F\'elix Viallet, 38031 Grenoble
Cedex, France; myriam.preissmann@g-scop.inpg.fr}}

\maketitle

\begin{abstract}
For integers $m,k \geq  1$, we investigate the maximum size of a
directed cut in directed graphs in which there are $m$ edges and each
vertex has either indegree at most $k$ or outdegree at most $k$.
\end{abstract}

\section{Introduction}

We deal with directed graphs, called here {\it digraphs}, without
loops and parallel edges.  An edge $xy$ of a digraph is interpreted as
an arc or an {\it arrow} going from the starting vertex or {\it tail}
$x$ to the end vertex or {\it head} $y$.  The {\it indegree} and the
{\it outdegree} of a vertex $v\in V(D)$ is respectively defined as
$d_{D}^-(v)= |\{zv\in E(D)| z\in V(D)\}|$ and $d_{D}^+(v)= |\{vw\in
E(D)| w\in V(D)\}|$.

Let $X,Y$ be a partition of the vertex set $V(D)$ of a digraph $D$.
The edge set $\{xy\in E(D)| x\in X, y\in Y\}$ is called a {\it
directed cut}.  Clearly a directed cut of a digraph $D$ does not
contain a directed path on three vertices (a $P_{3}$).  On the other
hand every directed $P_{3}$-free subgraph of $D$ is the subgraph of
some directed cut.  Thus when estimating the size of maximum directed
cuts we must find directed $P_{3}$-free subgraphs as large as
possible.  The {\it size of a cut} is its cardinality, the {\it size
of a digraph} is the cardinality of its edge set.

Discussions in \cite{alonetal} show that a digraph $D$ of size $m$ has
a cut of size $\frac{1}{4}m + \Theta(m^{1/2})$.  Furthermore, if the
outdegree of each vertex of $D$ is at most $k$, then $D$ has a cut of
size at least $(\frac{1}{4}+\frac{1}{8k+4})m$.  In \cite{cropperetal},
lower bounds for the largest directed cuts were asked for a family of
digraphs with constrained indegree or outdegree.  Let $D(k,\ell)$ be
the family of all digraphs in which every vertex has either indegree
at most $k$ or outdegree at most $\ell$ (that is $d^{-}(v)\leq k$ or
$d^{+}(v)\leq \ell$, for all $v\in V(D)$).  Note that a directed cut
(or any directed $P_{3}$-free graph) forms a graph that belongs to
$D(0,0)$.

In Section~\ref{secd11} we consider the case $k= \ell= 1$ and discuss
the size of the maximum directed cut of digraphs in $D(1,1)$.  It was
proved in \cite{alonetal} that every acyclic digraph of size $m$ in
$D(1,1)$ has a directed cut of at least $2m/5$ edges.  From a result
of Bondy and Locke \cite{bondyetal} it is easy to see that the same
lower bound holds for maximum directed cuts in triangle-free subcubic
digraphs (a graph is \emph{subcubic} if it has maximum degree at most
three).  Our main result in Theorem~\ref{cutformula} is the extension
of this bound for all digraphs in $D(1,1)$ as follows: if $D$ contains
at most $t$ pairwise disjoint directed triangles, then $D$ has a
directed cut of size at least $(2m-t)/5$.  The proof yields a
polynomial algorithm which actually finds a directed cut of that size
(Corollary \ref{alg}).

Theorem~\ref{cutformula} implies that
every digraph of size $m$ in $D(1,1)$ has a directed cut with at least
$m/3$ edges (a result first proved in \cite{alonetal}).  Furthermore,
every {\it connected} digraph of size $m$ in $D(1,1)$ has a directed
cut with at least $7m/20$ edges (see Theorem~\ref{7/20}).

In Section~\ref{secdkk} we consider digraphs in $D(k,k)$ for any $k$.
First we prove a decomposition property in Theorem \ref{decomp}: the
edge set of every digraph in $ D(p_1+p_2, p_1+p_2)$ can be partitioned
into two subgraphs one in $D(p_1, p_1)$ and the other in $D(p_2,
p_2)$.  In Theorem~\ref{k-1} we prove the lower bound $(2k-1)m/(2k+1)$
on the maximum size of a subgraph of $D$ belonging to $D(k-1,k-1)$.
It is worth noting that the regular tournament on $2k+1$ vertices has
no subgraph in $D(k-1,k-1)$ with more than $(2k-1/k)m/(2k+1)$ edges.

In Section~\ref{secd22} we show that if $D\in D(k,k)$ is acyclic and
has $m$ edges, then it contains a directed cut of size at least
$(\frac{1}{4}+\frac{1}{8k+4})m$ (see Theorem~\ref{acyclic}).  It is
worth noting that in a digraph $D\in D(k,k)$ one cannot guarantee a
directed cut of size larger than that proportion.  This is shown by
the regular tournament on $2k+1$ vertices, which has no directed cut
of size more than $(\frac{1}{4}+\frac{1}{8k+4}){2k+1\choose 2}$ (see
in \cite{alonetal}).  For $k= 2$ this ratio is $3m/10$.  In
Theorem~\ref{3/10} we can show that actually every digraph $D\in D(2,
2)$ with $m$ edges has a directed cut of size at least $3m/10$.  In
the proof of Theorems~\ref{acyclic} and~\ref{3/10} we use an
elementary counting method similar to those applied in
\cite{alonetal}.

Section~\ref{problems} concludes with open problems for further
consideration.  A challenging question whose answer we would like to
see the most is whether Theorem~\ref{acyclic} remains true for all
digraphs in $D(k,k)$, and for every $k\geq  3$.

\section{Maximum directed cut of digraphs in $D(1,1)$}
\label{secd11}

It was proved in \cite{alonetal} that every acyclic digraph of size
$m$ in $D(1,1)$ has a directed cut of at least $2m/5$ edges.  This is
not true for all digraphs in $D(1,1)$.  For example the directed
triangle, which is a member of $D(1,1)$, has no directed cut with two
edges.  Hence there are digraphs of size $m$ with maximum directed cut
not larger than $m/3$.  On the other hand, it was shown in
\cite{alonetal} that the edge set of every digraph $D\in D(1,1)$ has a
decomposition into three directed cuts (see Theorem~\ref{three}
below), hence $D$ always contains a directed cut of size $m/3$.

One might conclude that the ratio $m/3$ cannot be improved to $2m/5$
in general, and the graph that consists of disjoint directed triangles
is an obvious example showing that.  Actually the following example
shows that, for infinitely many values of $m$, there are even connected
digraphs in $D(1,1)$ of size $m$, that contain no cut of size
$3m/8$.

\

\noindent {\it Example 1.} For $i= 1, \dots, k$ let $H_i$ be a
directed path with five vertices $(u_i, v_i, w_i, x_i, y_i)$ plus the
chord $v_ix_i$, such that the $H_i$'s are pairwise disjoint.  Add
$k+1$ directed triangles $(y_i, u_{i+1}, z_i)$, for $i= 0, \dots, k$,
where $y_0, u_{k+1}$ and $z_0, z_1, \dots, z_k$ are distinct new
vertices.  The obtained graph $H$ has $m= 8k+3$ edges and its maximum
directed cut has size $3k+1= (3m-1)/8$.  

\ 

In spite of the evidence that the maximum directed cut size to edge
count ratio $2/5$ cannot be achieved, we show in the next theorem that
$1/3$ improves to $2/5$, in some sense, for all digraphs in $D(1, 1)$.

\begin{theorem}
\label{cutformula}
Let $D$ be a digraph in $D(1,1)$ with $m$ edges, and let $t$ be the
maximum number of pairwise disjoint directed triangles in $D$.  Then
$D$ has a directed cut of size at least $(2m-t)/5$.
\end{theorem}
\begin{proof}
The claim is clearly true for $m\leq 3$ and $t= 0$.  If $m= 3$ and
$t=1$, then $D$ is the directed triangle and any edge of the triangle
forms a directed cut of size $1= (2m-t)/5$.  Now let $D$ be a
counterexample with $m\geq 4$ edges, and assume that the theorem is
true for all digraphs in $D(1, 1)$ with at most $m-1$ edges.  Clearly
$D$ is connected.

Let $D^{+}$ be the subgraph of $D$ induced by $V^+= \{v\in V(D)|
d^{+}(v)\geq 2\}$; and let $D^{-}$ be the subgraph of $D$ induced by
$V^-= \{v\in V(D)| d^{-}(v)\geq 2\}$.  Notice that $v\in V^+$ implies
that $d^{-}(v)\leq 1$ and $v\in V^-$ implies $d^{+}(v)\leq 1$.

Because $D\in D(1, 1)$, if two directed triangles of $D$ have a common
vertex, then they must share a common edge.  Moreover, if a triangle
intersects with at least two other triangles, then they all share the
same common edge.  The following property of triangles will be useful.
\begin{claim}
     \label{cl11}
Every directed triangle of $D$ is contained in $D^+$ or in $D^-$.
\end{claim}
Assume that $T= (x, y, z)$ is a directed triangle with $d^-(x)= 1$ and
$d^+(y)= 1$.  Remove the edges of $T$ from $D$.  The graph $D'$ that
remains has $m' = m-3$ edges, and the maximum number $t'$ of disjoint
triangles in $D'$ satisfies $t'\leq t-1$.  By induction, $D'$ contains
a directed cut $K$ of size at least $ (2m' - t')/5\geq (2m-t)/5 -1$.
Obviously, $K\cup\{xy\}$ is still a directed $P_{3}$-free subgraph of
$D$ containing $ (2m-t)/5$ edges, a contradiction.  Therefore, either
$d^+(w)\geq 2$ for all $w\in \{x, y, z\}$ or $d^-(w)\geq 2$ for all
$w\in \{x, y, z\}$.  In the first case $T\subset D^+$ and in the
second case $T\subset D^-$.  Thus Claim~\ref{cl11} holds.  

\

Let $A, B\subset E(D)$ be a pair of disjoint edge sets such that every
directed $P_{3}$ in $D$ that has one edge in $A$ has its second edge
in $B$.  Note that this implies that $A$ contains no directed $P_{3}$.
We call any such pair $A, B\subset E(D)$ a {\it reducing pair}.  It is
clear that if $K$ is any directed cut in the digraph $D\setminus(A\cup
B)$, then $K\cup A$ is a directed $P_{3}$-free subgraph of $D$.  The
following claim will be used several times in the induction step.
\begin{claim}
\label{cl12}
$D$ has no reducing pair $A, B\subset E(D)$ with $|B|\leq
\frac{3}{2}|A|$.
\end{claim}
Suppose that $A, B\subset E(D)$ is a reducing pair with $|B|\leq
\frac{3}{2}|A|$.  Let $K$ be a largest directed cut in the digraph
$D'= D\setminus(A\cup B)$.  Then $K\cup A$ is a directed $P_{3}$-free
subgraph of $D$.  Digraph $D'$ has $m'= m-|A\cup B|$ edges, hence it
follows by induction that $|K|\geq (2m' - t)/5$.  We obtain $$|K\cup
A|\geq \frac{2m'-t}{5} + | A| \geq \frac{2m-t}{5}-\frac{2}{5}(|A|+|
B|)+|A|\geq \frac{2m-t}{5},$$
which contradicts the assumption that $D$ is a counterexample to the
theorem.  Thus Claim~\ref{cl12} holds.

\begin{claim}
     \label{cl13}
Each of $D^{+}$ and $D^{-}$ is a disjoint union of directed cycles.
Furthermore, every vertex in $D^{+}$ or $D^{-}$ is incident with
exactly one edge of $D\setminus(D^{+}\cup D^{-})$.
\end{claim}
Let $C$ be any connected component of $D^{-}$.  We show that $C$ is a
directed cycle.  By the definition of $D^{-}$, $C$ is either a rooted
tree with all edges directed towards the root, or a {\it function
graph} which is a rooted tree plus an edge from the root to some
vertex of the tree.

If $C$ is not a directed cycle, then it is either a singleton vertex
$v_{0}$ or it has a leaf $v_{0}$.  In each case, because $v_{0}$ is in
$V^-$, there exist distinct edges $e_{1}= v_{1}v_{0}, e_{2}=
v_{2}v_{0}$ of $D$.  Furthermore, $d^{+}(v_{0})\leq 1$, thus at most
one edge $f_{0}$ leaves $v_{0}$.  Since $v_{1}, v_{2}$ are not in $C$,
they are not in $V^-$, hence at most one edge enters each, say $f_{1}$
and $f_{2}$, respectively.  Note that both edges exist in $A= \{e_{1},
e_{2}\}$, but any edge from the set $B= \{f_{0}, f_{1}, f_{2}\}$ might
actually not exist.  In either case, $A, B$ form a reducing pair with
$|B|\leq \frac{3}{2}|A|$, contradicting Claim~\ref{cl12}.  Thus every
component of $D^-$ is a directed cycle.  Furthermore, if there are two
edges $e_{1}, e_{2}$ of $D\setminus(D^{+}\cup D^{-})$ at some vertex
$v_{0}\in V^-$, then one obtains a contradiction using the same
reducing pair.

An analogous argument shows that every connected component $C$ of
$D^{+}$ is a directed cycle with exactly one edge of
$D\setminus(D^{+}\cup D^{-})$ at each vertex of $C$.  Thus
Claim~\ref{cl13} holds.  

\

Note that, due to Claims~\ref{cl11} and~\ref{cl13}, all directed
triangles of $D$ are among the cycles of $D^{+}$ and $D^{-}$.

\begin{claim}
     \label{cl14}
All directed cycles in $D^{+}$ and $D^{-}$ have odd length.
\end{claim}
Suppose the contrary, and let $C= (x_{1}, x_{2}, \dots, x_{2p})$ be a
directed cycle, say in $D^{+}$.  For every $i= 1, \dots, p$ let
$e_{2i-1}, e_{2i}$ be the two edges going out from $x_{2i}$, such that
$e_{2i}= x_{2i}x_{2i+1}$, and call $y_{i}$ the end vertex of
$e_{2i-1}$.  Then $y_{i}\in V\setminus V^{+}$, therefore there is at
most one edge $g_{i}$ going out from $y_{i}$.  For every $i= 1, \dots,
p$, let $f_{2i-1}, f_{2i}$ be the two edges going out from $x_{2i-1}$,
such that $f_{2i-1}= x_{2i-1}x_{2i}$.  Let $A= \{e_{1}, \dots,
e_{2p}\}$ and $B= \{f_{1}, \dots, f_{2p}\}\cup \{g_{1}, \dots,
g_{p}\}$.  Observe that $A, B$ are disjoint and that $B$ contains one
edge of each directed $P_{3}$ of $D$ that has an edge in $A$.
Therefore $A, B$ is a reducing pair, with $|B|\leq \frac{3}{2}|A|$,
contradicting Claim~\ref{cl12}.  Thus Claim~\ref{cl14} holds.

\begin{claim}
     \label{cl15}
     $D^{+}$ and $D^{-}$ have the same number of vertices, say this
     number is $k$, and $D\setminus(D^{+}\cup D^{-})$ is the union of
     $k$ disjoint edges going from $D^{+}$ to $D^{-}$.
\end{claim}
We shall prove that $V^{0}= V(D)\setminus (V^-\cup V^{+})= \emptyset$.
Assume on the contrary that $V^{0}\neq\emptyset$.  By the connectivity
of $D$, there is a vertex $y\in V^{0}$ adjacent to some vertex of
$D^{+}\cup D^{-}$.  By symmetry, we may assume that $yz$ is an edge
for some $z\in V^-$.  Let $C\subseteq D^{-}$ be the directed cycle
containing $z$, let $C$ have length $2\ell+1$, with $\ell\geq 1$.  We
call \emph{$(\ell+1)$-set} any subset $L\subset V(C)$ such that
$V(C)\setminus L$ is a maximum independent set of $C$.  Note that for
any two vertices $x,y$ of $C$ there exists an $(\ell+1)$-set that
contains both $x,y$.

If $d^-(y)\neq 0$, then let $e_{0}= xy$, and let $g_{0}$ be an edge
going into $x$ if it exists.  (Note that $y\notin V^{-}$ implies
$x\notin V^{-}$.)  Let $L\subset V(C)$ be an $(\ell+1)$-set of $C$
not containing $z$ and define:
  \begin{eqnarray*}
B_{1 }&=& \{f\in E(C) \mid f= ww' \;\hbox{for some}\; w\in L\}, \\
A &=& \{ e_{0}\}\cup (E(C)\setminus B_{1})\cup \{e\notin E(C) \mid e=
vw \mbox{ for some } w\in L\}, \\
B_{2} &=& \{g\in (E(D)\setminus B_{1})\mid g= uw \mbox{ such that }
wv\in A\}.
  \end{eqnarray*}
Observe that $A$ contains no directed $P_{3}$ and that every directed
$P_3$ with one edge in $A$ has its other edge in $B=B_1\cup B_2$.  So
$A,B$ is a reducing pair.  Since $|A|= 2\ell +2$, $|B_{1}|= \ell +1$,
and $|B_{2}|\leq 2\ell +2$, we have $|B|\leq \frac{3}{2}|A|$,
contradicting Claim~\ref{cl12}.

If $d^-(y) = 0$, then let $L\subset V(C)$ be an $(\ell+1)$-set of $C$
containing $z$, and define:
  \begin{eqnarray*}
  B_{1} &=& \{f\in E(C) \mid f= ww' \mbox{ for some } w\in L\}, \\
  A &=& (E(C)\setminus B_{1})\cup \{e\notin E(C) \mid e= vw \mbox{ for
  some } w\in L\}, \\
  B_{2} &=& \{g\in (E(D)\setminus B_{1}) \mid g= uw \mbox{ such that }
  wv\in A\}.
  \end{eqnarray*}
Again, $A$ and $B= B_{1}\cup B_{2}$ form a reducing pair.  We have
$|A|= 2\ell +1$, $|B_{1}|= \ell +1$ and $|B_{2}|= 2\ell$ since no edge
enters into $y$.  Hence $|B|< \frac{3}{2}|A|$, contradicting
Claim~\ref{cl12}.  Then Claim~\ref{cl15} follows from the second part
of Claim~\ref{cl13}.

\ 

Call $M$ the (loopless) bipartite multigraph obtained by contracting
of every directed cycle into one vertex.
\begin{claim}
     \label{cl16}
$M$ is a simple graph.
\end{claim}
Suppose on the contrary that there are at least two edges from the
cycle $C^{+}\subseteq D^{+}$ to the cycle $C^-\subseteq D^{-}$.  By
Claim~\ref{cl15}, $C^{+}$ is an odd cycle, thus there exist edges $ux,
vy\in E(D)$ with $u, v\in V(C^{+})$, $x, y\in V(C^-)$ such that $(u,
b_{1}, \dots, b_{2q}, v)$ is a directed subpath of $C^{+}$, and no
vertex $b_{i}$ has an edge to $C^-$ ($q= 0$ means that $uv$ is an edge
of $C^{+}$).

Let $C^-$ have length $2\ell+1$ (it is an odd cycle by
Claim~\ref{cl15}).  Obviously there exists an $(\ell+1)$-set $L\subset
V(C^-)$ including $x$ and excluding $y$.  Define:
  \begin{eqnarray*}
   B_{1} &=& \{f\in E(C^-) \mid f= wz \mbox{ for some } w\in L\}, \\
   A_{0} &=& (E(C^-)\setminus B_{1})\cup \{e\notin E(C^-) \mid e= zw
   \mbox{ for some } w\in L\}, \\
   B_{2}&=& \{g\in (E(D)\setminus B_{1}) \mid g= wz \mbox{ such that }
   zw'\in A_{0}\}.
  \end{eqnarray*}
Let $e'_{0}= ub_{1}$, $f'_{0}= vw$ where $w\in V(C^{+})$, $g'_{0}=
vy$, and define:
  \begin{eqnarray*}
   A_{0}' &=& \{e'_{0}\}\cup \{e'\in E(D) \mid e'= b_{2i}z,\; 1\leq
   i\leq q\}, \\
   B_{1}' &=& \{f'_{0}\}\cup \{f'\in E(D) \mid f'= b_{2i-1}z,\; 1\leq
   i\leq q\}, \\
   B_{2}' &=& \{g'\in (E(D)\setminus B_{1}') \mid g'= zw' \mbox{ such
   that } bz\in A_{0}'\}\setminus\{g'_{0}\}.
  \end{eqnarray*}
Observe that the set $A= A_{0}\cup A_{0}'$ contains no directed
$P_{3}$, and every directed $P_{3}$ with an edge in $A$ has its other
edge in $B= B_{1}\cup B_{2}\cup B_{1}'\cup B_{2}'$ from $D$.  Hence
$A, B$ form a reducing pair.  We have $|A_{0}|= 2\ell +1$, $|B_{1}|=
\ell +1$, $|B_{2}|= 2\ell +1$, $|A_{0}'|= 2q+1$, $|B_{1}'|= 2q+1$,
$|B_{2}'|= q$, so $|A|= 2(\ell+q+1)$ and $|B| = 3(\ell+q+1)
=\frac{3}{2}|A|$, contradicting Claim~\ref{cl12}.  Thus
Claim~\ref{cl16} holds.

\

Because every vertex of the contraction graph $M$ has degree at least
three, $M$ has a cycle.  To conclude the proof of the theorem we show
that this leads to a contradiction.

\

Consider a shortest cycle $\gamma\subset M$, and let $\gamma=
(C^{+}_{1}, C^{-}_{1}, C^{+}_{2}, C^{-}_{2}, \dots, C^{+}_{p},
C^{-}_{p})$, where, for each $i\in\{1, \ldots, p\}$,
$C^{+}_{i}\subseteq D^{+}$ and $C^{-}_{i}\subseteq D^{-}$ are cycles
of $D$ of odd length.  The edges of $\gamma$ correspond to a matching
of $D$ from the set $\cup_{i= 1}^q\{u^{i}, v^{i}\}$, to the set
$\cup_{i= 1}^q\{x^{i}, y^{i}\}$, where $u^{i}, v^{i}\in C^{+}_{i}$ and
$x^{i}, y^{i}\in C^{-}_{i}$.  Furthermore, by Claim~\ref{cl16} and
since $\gamma$ has no chords in $M$, no more edges of $D$ are induced
between these cycles.  We may assume, so we do, that $(u^{i}, b^i_{1},
\dots, b^i_{2q_{i}-1}, v^i)$, where $q_{i}\geq 1$, is a directed
subpath of $C^{+}_{i}$.

Let $2\ell_{i}+1$ be the length of $C^{-}_{i}$.  For every $i= 1,
\dots, p$ select an $(\ell_{i}+1)$-set $L_{i}\subset V(C^{-}_{i})$ of
$C^{-}$ such that $x^i, y^i\in L_{i}$, and define the following sets:
  \begin{eqnarray*}
   B^{1}_{i} &=& \{f\in E(C^{-}_{i}) \mid f= wz \mbox{ for some } w\in
   L_{i}\}, \\
   A^{1}_{i} &=& (E(C^{-}_{i})\setminus B^{1}_{i})\cup \{e\notin
   E(C^{-}_{i}) \mid e= zw \mbox{ for some } w\in L_{i}\}, \\
   B^{2}_{i} &=& \{g\in (E(D)\setminus B^{1}_{i}) \mid g= uz \mbox{
   such that } zw\in A^{1}_{i}\}.\\
  \end{eqnarray*}
Let $A^{1}= \cup_{i= 1}^p A^{1}_{i}$ and $B^{1}= \cup_{i=1}^p
(B^{1}_{i}\cup B^{2}_{i})$.  We have $|A^{1}|= \sum_{i=1}^p
|A^{1}_{i}|= \sum_{i= 1}^p(2\ell_{i} +1)$, and because $|B^{1}_{i}|=
\ell_{i} +1, |B^{2}_{i}|= 2\ell_{i} +1$, we obtain $|B^{1}|=
\sum_{i=1}^p (3\ell_{i} +2)$.

For every $i= 1, \dots, p$, let $e_{i}= u^ib^i_{1}$, $f_{i}=
b^i_{2q_{i}-1}v^i$, and define sets:
  \begin{eqnarray*}
   A^{2}_{i} &=& \{e_{i}\}\cup \{e\in (E(D) \mid e= b_{2j}w,\; 1\leq
   j\leq q_{i}-1\}, \\
   B^{3}_{i} &=& \{f\in E(D) \mid f= b_{2j-1}w,\; 1\leq j\leq
   q_{i}\}\setminus \{f_{i}\}, \\
   B^{4}_{i} &=& \{g\in (E(D)\setminus B^{3}_{i}) \mid g= wz \mbox{
   such that } bw\in A^{2}_{i}\}\setminus \{f_{i}\}.\\
  \end{eqnarray*}
Let $A^{2}= \cup_{i= 1}^p A^{2}_{i}$ and $B^{3}= \cup_{i= 1}^p
(B^{3}_{i}\cup B^{4}_{i})$.  We have $|A^{2}|= \sum_{i=1}^p
|A^{2}_{i}|= \sum_{i=1}^p (2q_{i} -1)$, and because $|B^{3}_{i}|=
2q_{i} -1$ and $|B^{4}_{i}|= q_{i} -1$, we obtain $|B^{3}|=
\sum_{i=1}^p (3q_{i} -2)$.  Observe that the sets $A= A^{1}\cup A^{2}$
and $B= B^{1}\cup B^{3}$ form a reducing pair.  Furthermore, $|A|=
\sum_{i= 1}^p(2\ell_{i} + 1 + 2q_{i} -1)= 2\sum_{i= 1}^p(\ell_{i} +
q_{i})$ and $|B|= \sum_{i= 1}^p(3\ell_{i} + 2 +3q_{i} -2)= 3\sum_{i=
1}^p(\ell_{i} + q_{i})= \frac{3}{2}|A|$, contradicting
Claim~\ref{cl12}.  This concludes the proof of the theorem.
\end{proof}

The proof of the theorem can be formulated as an algorithm which,
given any digraph $D\in D(1,1)$ with $m$ edges and at most $t$
disjoint directed triangles, constructs a directed cut $K$ of size at
least $(2m-t)/5$.  We sketch such an algorithm here.  Start from
$K:=\emptyset$.  Then apply the following general step.  Find the
subgraphs $D^+$ and $D^-$.  If there is a directed triangle that is
not included in $D^+$ or $D^-$, with the notation of Claim~\ref{cl11},
then set $K:=K\cup \{xy\}$ and iterate with the subgraph $D\setminus
\{xy, yz, zx\}$.  (When iterating, the subgraphs $D^+, D^-$ must be
updated.)  If there is no such directed triangle, then either $D$
violates one of Claims~\ref{cl13}--\ref{cl16}, or $D$ satisfies the
conditions described after the proof of Claim~\ref{cl16}; and in
either case, the proof of the theorem shows how to find a reducing
pair $(A, B)$.  Then set $K:=K\cup A$ and iterate the general step
with the subgraph $D\setminus (A\cup B)$.  The algorithm terminates
when $D$ becomes edgeless.  Then at termination $K$ is a directed cut
of size at least $(2m-t)/5$.  It is easy to see that all the
operations (updating $D^+$ and $D^-$, finding a directed triangle,
checking whether $D$ violates one of the claims, determining the
structure described after the proof of Claim~\ref{cl16}) can be done
in polynomial time, and there are at most $m$ iterations.  Thus we
obtain:
\begin{corollary}
      \label{alg}
There is a polynomial time algorithm which, given any digraph $D\in
D(1,1)$ with $m$ edges and at most $t$ disjoint directed triangles,
finds a directed cut in $D$ of size at least $(2m-t)/5$.\hfill $\Box$
  \end{corollary}
%
\begin{corollary}
      \label{m/3}
If $D\in D(1, 1)$ has $m$ edges, then it contains a directed cut of
size at least $m/3$.  Moreover $D$ has no directed cut of size larger
than $m/3$ if and only if $D$ is the union of disjoint directed
triangles.
  \end{corollary}
\begin{proof}
The number of pairwise disjoint directed triangles satisfies $t\leq
m/3$, with equality if and only if $D$ is a union of disjoint directed
triangles.  Now the claim follows by Theorem~\ref{cutformula}, because
$(2m-t)/5\geq (2m-m/3)/5= m/3$.
\end{proof}

\begin{corollary}
\label{notriangle}
If $D\in D(1, 1)$ has $m$ edges and no directed triangle, then it
contains a directed cut of size at least $2m/5$.\hfill $\Box$
\end{corollary}

Results by Bondy and Locke \cite{bondyetal} on the bipartite density
of (undirected) subcubic graphs are reminescent of our investigations
concerning $D(1,1)$.  They proved in \cite{bondyetal} that a
triangle-free subcubic graph has a bipartite subgraph of size at least
$4m/5$.  Observe that any triangle-free digraph of maximum degree at
most three belongs to $D(1,1)$, and it is obtained from a subcubic
graph by orienting its edges.  Hence their result implies that such a
$D$ has a directed cut of size at least $2m/5$, the half of $4m/5$.
Corollary~\ref{notriangle} shows that this bound is valid for the much
larger class of digraphs in $D(1,1)$ containing no directed triangle.

Now we show that the lower bound $m/3$ in Corollary~\ref{m/3} can be
surpassed for connected digraphs of $D(1, 1)$.
\begin{theorem}
      \label{7/20}
If $D\in D(1, 1)$ is a connected digraph with $m$ edges, and $D$ is
not a triangle, then it contains a directed cut of size at least
$7m/20$.
  \end{theorem}
\begin{proof}
The proof works by induction on $m$.  Let $t$ be the maximum number of
pairwise disjoint directed triangles of $D$.  By the hypothesis, we
have $t=0$ if $m\le 3$ and $t\leq 1$ if $m= 4, 5, 6$.  Thus by
Theorem~\ref{cutformula}, there is a cut of size at least $1, 1, 2, 2,
2, 3$, respectively, for $m= 1, \ldots, 6$, which matches the
corresponding value of $\lceil 7m/20\rceil$.  Now let $m\geq 7$, and
assume that the claim is true for connected graphs with strictly less
than $m$ edges.  Observe that any $t$ disjoint directed triangles of
$D$ have a total of $3t$ edges, furthermore, by the connectivity of
$D$, there are at least $t-1$ more edges between these triangles.
Hence we have $m\geq 4t-1$.

If $m> 4t-1$, or equivalently, if $t\leq m/4$, then by
Theorem~\ref{cutformula}, $D$ has a cut of size at least $(2m-t)/5\geq
(2m-m/4)/5 = 7m/20$ edges as stated.

Assume now that $m= 4t-1$.  So $D$ consists of $t$ disjoint directed
triangles connected by $t-1$ edges in a tree-like manner.  Since we
cannot have $t=1$ and $m=3$, we have $t\ge 2$.  So there is a directed
triangle $T= (x, y, z)$ that is adjacent to exactly one edge, say
$xx'$, which is adjacent to another directed triangle $T'= (x', y',
z')$.  (The symmetric argument applies if the orientation of the edge
between $T$ and $T'$ is $x' x$.)  Removing from $D$ the vertices and
edges of $T$ together with the two edges $xx', x' y'$, we obtain a
connected digraph $D'$ with $m'= m-5\geq 2$ edges.  By the induction
hypothesis, $D'$ has a cut $H'$ of size at least $7m'/20= (7m-35)/20 >
7m/20 -2$ edges.  Clearly $H'\cup\{xx', yz\}$ has no directed $P_{3}$,
which yields a cut of size at least $7m/20$ in $D$.
\end{proof}

Just like with Theorem~\ref{cutformula}, the proof of
Theorem~\ref{7/20} can be formulated easily as a polynomial time
algorithm (we omit the details).  So we have:
\begin{corollary}
      \label{alg720}
There is a polynomial time algorithm which, given any digraph $D\in
D(1,1)$ with $m$ edges, such that no component of $D$ is a directed
triangle, finds a directed cut in $D$ of size at least $7m/20$.\hfill
$\Box$
\end{corollary}

\section{Decompositions of $D(k, k)$}
\label{secdkk}

The problem of covering the edges of a digraph with cuts was proposed
in \cite{alonetal}.  Upper bounds were given for digraphs in $D(k,
\ell)$, and the only exact value was determined for $k= \ell= 1$.

\begin{theorem}[\cite{alonetal}]
  \label{three}
The edge set of any digraph $D\in D(1, 1)$ can be decomposed into at
most three cuts.\hfill $\Box$
\end{theorem}

\begin{theorem}
     \label{decomp}
For integers $p_1, p_2\ge 0$, the edge set of every digraph $D\in
D(p_1+p_2, p_1+p_2)$ can be decomposed into two subgraphs $D_1\in
D(p_1, p_1)$ and $D_2\in D(p_2, p_2)$.
\end{theorem}
\begin{proof}
Since $D$ is in $D(p_1+p_2, p_1+p_2)$, its vertex set $V(D)$ can be
partitioned into two sets $X, Y$ such that every vertex $x\in X$
satisfies $d^-(x)\le p_1+p_2$ and every vertex $y\in Y$ satisfies
$d^+(y)\le p_1+p_2$.  Consider the set of edges $B= \{yx\in E(D) |
y\in Y, x\in X\}$.  By the definition of $X$ and $Y$, in the bipartite
graph $(X, Y; B)$ every vertex has degree at most $p_1+p_2$.  By a
classical corollary of the K\H{o}nig-Hall theorem (see e.g.,
\cite[Prop.~5.3.1]{die}), the edges of $B$ can be colored with
$p_1+p_2$ colors so that any two adjacent edges have different colors.
Let $B_1$ be the set of edges of $B$ with the first $p_1$ colors and
$B_2$ be the set of edges of $B$ with the remaining colors.

For every vertex $x\in X$, the set $E^-(x)$ of edges with end $x$ has
size at most $p_1+p_2$, so it can be partitioned into two sets
$E_1(x)$ and $E_2(x)$ such that, for $j= 1, 2$, $|E_j(x)|\le p_j$ and
$E^-(x)\cap B_j\subseteq E_j(x)$.  Likewise, for every vertex $y\in
Y$, the set $E^+(y)$ of edges with origin $y$ has size at most
$p_1+p_2$, so it can be partitioned into two sets $E_1(y)$ and
$E_2(y)$ such that, for $j= 1, 2$, $|E_j(y)|\le p_j$ and $E^+(y)\cap
B_j\subseteq E_j(y)$.

Finally let the set $\{xy\in E(D)| x\in X, y\in Y\}$ be partitioned
arbitrarily into two sets $F_1, F_2$.  Now, for $j= 1, 2$, let $D_j$
be the subgraph of $D$ whose edge set is $B_j\cup F_j\cup
\bigcup_{x\in V} E_j(x)$.  The definition of these sets implies that
each edge of $D$ lies in exactly one of $D_1, D_2$ and that $D_j\in
D(p_j, p_j)$ for $j= 1, 2$.  More precisely, for $j= 1, 2$, in $D_j$
every vertex $x\in X$ satisfies $d^-(x)\le p_j$ and every vertex $y\in
Y$ satisfies $d^+(y)\le p_j$.
\end{proof}

\begin{corollary}
     \label{twotwo}
The edges of every digraph $D\in D(2, 2)$ can be decomposed into two
subgraphs $D_1, D_2\in D(1, 1)$.  \hfill$\Box$
\end{corollary}

 From Corollary~\ref{twotwo} and Theorem~\ref{three} it follows that
every digraph $D\in D(2, 2)$ can be covered with six directed cuts.
If there was a decomposition of $D$ into a cut and a digraph in $D(1,
1)$, then $D$ would have a cut cover only with four cuts, by Theorem
\ref{three} again.  Our next example shows that such a decomposition
is not always possible.

\

\noindent {\it Example 2.} Take two disjoint copies of a regular
tournament on five vertices, $G_{1}$, $G_{2}$, and include all $25$
edges directed from $G_{1}$ to $G_{2}$.  Thus we obtain a digraph
$H\in D(2, 2)$.  Assume that $K\subset E(H)$ is a cut such that $H'=
H\setminus K$ is in $D(1, 1)$.  The regular tournament has no cut with
more than three edges, hence $G_{1}$ has a vertex $v_{0}$ such that
every edge going into $v_0$ is in $E(H)\setminus K$ and at least one
edge going out of $v_0$ is in $E(H)\setminus K$.  Thus
$d^-_{H'}(v_{0})= 2$, which implies that $v_{0}z\in K$ for all $z\in
V(G_{2})$ in order to obtain $d^+_{H'}(v_{0})\leq 1$.  Then it follows
that no edge of $G_{2}$ belongs to $K$, thus $d^+_{H'}(z)= 2$ and
$d^-_{H'}(z)\geq 2$ for all $z\in V(G_{2})$, a contradiction.

\

How large a subgraph belonging to $D(1, 1)$ can be found in a digraph
$D\in D(2, 2)$?  Corollary~\ref{twotwo} implies that $D$ with $m$
edges contains a subgraph in $D(1, 1)$ with at least $m/2$ edges.  A
larger bound will follow from our more general result.

\begin{theorem}
     \label{k-1}
Every digraph $D\in D(k, k)$ with $m$ edges has a subgraph belonging
to $D(k-1, k-1)$ with at least $(2k-1)m/(2k+1)$ edges.
   \end{theorem}
\begin{proof}
Let $W= \{v\in V(D)| d_{D}^-(v) \leq k\}$ and $B= \{v\in V(D)|
d_{D}^+(v) \leq k \}$.  Because $D\in D(k, k)$, we have $V(D)= W\cup
B$.  We say that $v\in W$ is \emph{white}, and $v\in B$ is
\emph{black}; note that a vertex may have both colors.  An edge $xy\in
E(D)$ is called a {\it black tail arrow} if $x\in B$, and it is called
a {\it white head arrow} if $y\in W$.  Note that an edge can be both a
black tail and a white head arrow.  Observe the symmetry of the colors
with respect to reversing all arrows in $D$.  Due to this symmetry, if
a property is verified for white vertices, then the analogous property
is true for black vertices with directions reversed.

Let $R\subset E(D)$ be a set of edges such that (a) the graph $D'=
D\setminus R$ is in $D(k-1, k-1)$, (b) $R$ is minimum among all sets
with property (a), and (c) $R$ has the maximum number of black tail
arrows and white head arrows (each arrow counted once) among all sets
that satisfy (a) and (b).  Clearly such a set $R$ exists.

For each edge $e=  xy\in R$, we define a {\it critical vertex} of $e$
as follows:

$x$ is a critical vertex for $e=  xy\in R$ if $d_{D'}^+(x)=  k-1$ and
$d_{D'}^-(x)\geq  k$;

$y$ is a critical vertex for $e=  xy\in R$ if $d_{D'}^-(y)=  k-1$ and
$d_{D'}^+(y)\geq  k$.

\noindent The minimality of $R$ means that at least one of $x,y$ is
critical for each edge $e= xy$ of $R$.  Note that both $x,y$ may be
critical for $e$.  For each $e\in R$, let $\crit(e)\subseteq\{x, y\}$
be the set of critical vertices of $e$.  For any subset $X\subseteq
R$, define $\crit(X)= \cup_{e\in X}\crit(e)$.  From here on, the word
critical vertex refers to elements of $\crit(R)$.  All critical
vertices are in the set $\{v\in V(D)| d_{D}^-(v), d_{D}^+(v) \geq k
\}$, however not every vertex in that set is critical for some edge of
$R$.  The main point of the proof is to establish that:
\begin{claim}\label{critr}
  $|\crit(R)|\geq  |R|$.
\end{claim}

Assume we already know that $|\crit(R)|\geq |R|$.  Then the definition
of critical vertices implies that, for every $v\in \crit(R)$, there
are at least $2k-1$ edges not in $R$ and incident with $v$.  Thus for
the size of $D$ we have the bound $m\geq |R| + (2k-1)|R|/2$, hence
$|R|\leq 2m/(2k+1)$.  So $D'$ has at least $m-|R|\geq (2k-1)m/(2k+1)$
edges, and the theorem follows.  Therefore the rest of the proof
consists in proving Claim~\ref{critr}.

\

The definition of critical vertices implies easily the following two
claims, whose proof is omitted.
\begin{claim}
\label{noncritical}
If $x\in B$ and $d_{R}^{+}(x)\geq  2$, then $x\notin \crit(R)$.  By
symmetry, If $x\in W$ and $d_{R}^{-}(x)\geq  2$, then $x\notin
\crit(R)$.
\end{claim}
\begin{claim}
\label{P3}
If $e= xy\in R$, $y\in B$, and $yz\in R$, then $\crit(e)= \{x\}$.  By
symmetry, if $e= xy\in R$, $x\in W$, and $ux\in R$, then $\crit(e)=
\{y\}$.
\end{claim}

Now we examine the subgraph formed by $R$.  Let $A\subseteq R$ be any
connected component of $R$ (we use $A$ and $R$ to denote the digraphs
defined by the edges in $A$ and $R$, respectively).  Note that
$\crit(R)= \crit(A)\cup \crit(R\setminus A)$.

\begin{claim}\label{atree}
If $A$ has no cycle, then $|\crit(A)|\geq  |V(A)|-1=  |A|$.
\end{claim}
In this case $A$ is a tree with $|A|+1$ vertices.  We show that at
most one non critical vertex may exist in $A$.  Suppose on the
contrary that $u, v$ are two non-critical vertices in $A$, and let $P=
(u, \dots, v)$ be the (unique) shortest chain between them in $A$.
Observe that $P$ has length at least $2$ (for otherwise its unique
edge $uv$ would satisfy $\crit(uv)= \emptyset$), and that the
inclusion into $D'$ of the two edges of $P$ incident to $u$ and $v$
does not increase their corresponding indegree or outdegree above
$k-1$.

For every white vertex $w$ of $V(P)\setminus\{u, v\}$ select a white
head arrow $xw$, and for every black vertex $z$ of $V(P)\setminus\{u,
v\}$ select a black tail arrow $zx$ (for two-colored vertices take one
such arrow arbitrarily).  Let $F$ be the set of selected arrows.  So
$|F|\le |P|-1$.  Define $R^{*}= (R\setminus P)\cup F$.  The graph
$D^{*}= D\setminus R^{*}$ belongs to $D(k-1, k-1)$, because the
outdegree of every black vertex of $V(P)\setminus\{u, v\}$, and the
indegree of every white vertex of $V(P)\setminus\{u, v\}$ is at most
$k-1$, furthermore the corresponding degrees of $u$ and $v$ do not
increase above $k-1$.  The set $R^{*}$ satisfies $|R^{*}|\le |R|-1$,
contradicting the minimality of $R$.  Thus $A$ has at most one
non-critical vertex, and Claim~\ref{atree} holds.

\

Now we consider an arbitrary cycle $C$ in $R$ (if any).

\begin{claim}
\label{cycle}
$C$ has no edge $e=  xy$ with $x\in W\setminus B$ and $y\in B\setminus
W$.
\end{claim}
Suppose that there is such an edge $e= xy$.  Note that $e$ is neither
a white head arrow nor a black tail arrow in $C$.  Hence $C$ has at
most $|C|-1$ white head and black tail arrows.  For every white vertex
$w\in V(C)$ select a white head arrow $zw$, and for every black vertex
$v\in V(C)$ select a black tail arrow $vu$ (for two-colored vertices
select one arrow arbitrarily).  Let $F$ be the set of $|C|$ selected
edges, and define $R^{*}= (R\setminus C)\cup F$.  The set $R^{*}$
satisfies $|R^{*}|\le |R|$, and contains more white head and black
tail arrows than $R$.  Furthermore, the graph $D^{*}= D\setminus
R^{*}$ belongs to $D(k-1, k-1)$, because the outdegree of every black
vertex of $C$, and the indegree of every white vertex of $C$ is at
most $k-1$.  This contradicts the choice of $R$.  Thus
Claim~\ref{cycle} holds.

\begin{claim}
\label{cl2}
Let $u, x, y, v$ be four consecutive vertices of $C$.  \\
(1) If $xu, xy, vy\in C$ and $x$ is black, then $y$ is not white.\\
(2) If $ux, xy, vy\in C$, then either $x$ or $y$ is not white.\\
(3) If $ux, xy, yv\in C$ and $y$ is black, then $x$ is not white.
\end{claim}
Suppose on the contrary that any of (1), (2), (3) fails.  Then, in
either case, the edge $e= xy$ satisfies $\crit(e)= \emptyset$, which
contradicts the minimality of $R$.  Thus Claim~\ref{cl2} holds.

\begin{claim}
     \label{cdm}
$C$ is a directed cycle and it is monochromatic, i.e., its vertices
are either all in $W\setminus B$ or all in $B\setminus W$.
\end{claim}
Suppose first that $C$ is not a directed cycle, and consider a longest
directed subpath $(x_{1}, \dots, x_{q})$ of $C$, where $q\geq 2$.

Suppose that $q= 2$, i.e., the directions of the edges alternate on
$C$.  Let $z_{1}w_{1}, z_{1}w_{2}, z_{2}w_{2}\in C$.  If $z_{1}\in
W\setminus B$, then by Claim~\ref{cycle} we have $w_{1}, w_{2}\in W$.
If $z_{1}\in B$, then Claim~\ref{cl2} (1) implies that $w_{2}\in
B\setminus W$, and $z_{2}\in B$ follows by Claim~\ref{cycle}.  Thus we
obtain that either $w_{1}, z_{1}, w_{2}\in W$ or (symmetrically)
$z_{1}, w_{2}, z_{2}\in B$.  We show a contradiction in the first
case, then, by symmetry, the second case is impossible as well.  So
assume that $w_{1}, z_{1}, w_{2}\in W$ and set $e_{i}= z_{1}w_{i}$,
$i= 1, 2$.  Select an arbitrary white head arrow $f= xz_{1}\in E(D)$.
The set $R^{*}= (R\setminus \{e_{1}, e_{2}\})\cup \{f\}$ satisfies
$|R^{*}|\le |R|-1$, and the graph $D^{*}= D\setminus R^{*}$ belongs to
$D(k-1, k-1)$, because $d_{D^*}^-(w_{i})= d_{D'}^-(w_{i})\leq k-1$ for
$i= 1, 2$, and $d_{D^*}^-(z_{1})\leq k-1$.  This contradicts the
minimality of $R$.  Therefore $q\ge 3$.

By Claim~\ref{cl2} (2), either $x_{q}$ or $x_{q-1}$ is not white on
the directed path $(x_{1}, x_{2}, \dots, x_{q})$, for $q\geq 3$.  If
$x_{q} \in B\setminus W$, then $x_{q-1}\in B$ by Claim~\ref{cycle}.
Thus that in each case $x_{q-1}$ is black.

Suppose that $q= 3$.  Let $e_{1}=  x_{1}x_{2}$, $e_{2}=  x_{1}y_{2}\in
C$, with $e_{1}\neq e_{2}$, and let $y_{3}$ be the second neighbor of
$y_{2}$ on $C$ different from $x_{1}$.

Assume first that $y_{2}y_{3}\in E(D)$.  Then, by the argument above,
$x_{2}$ and $y_{2}$ are both black.  Observe that $x_{1}\in W\setminus
B$, since otherwise the edges $e_{1}$ and $e_{2}$ have no critical
vertices.  Now select an arbitrary white head arrow $f=  zx_{1}\in
E(D)$.  The set $R^{*}=  (R\setminus \{e_{1}, e_{2}\})\cup \{f\}$
satisfies $|R^{*}|=  |R|-1$, and the graph $D^{*}=  D\setminus R^{*}$
belongs to $D(k-1, k-1)$, contradicting the minimality of $R$.

Assume now that $y_{3}y_{2}\in E(D)$ (where $y_{3}y_{2}$ might
coincide with $x_{2}x_{3}$, if $C$ is a triangle).  As before, we have
$x_{2}\in B$ and $x_{1}\in W\setminus B$.  Then, by Claim~\ref{cycle},
$y_{2}\in W$.  Selecting a white head arrow $f$ at $x_{1}$ and
defining the set $R^{*}=  (R\setminus \{e_{1}, e_{2}\})\cup \{f\}$ we
obtain a contradiction in the same way as before.  Therefore $q\ge 4$.

We already know that $x_{q-1}$ is black.  Hence $x_{q-2}\in B\setminus
W$ by Claim~\ref{cl2} (3).  Applying Claim~\ref{cl2} (3) repeatedly,
we obtain that $x_{q-2}, \dots, x_{2}$ are in $B\setminus W$.  Then we
have $x_{1}\in B$ by Claim~\ref{cycle}.  Hence the edge $x_{1}x_{2}$
has no critical vertex, because $d_{D'}^+(x_{1})\leq  k-2$ and
$d_{D'}^+(x_{2})\leq  k-1$, contradicting the minimality of $R$.  So we
have established that $C$ is a directed cycle.

Now assume without loss of generality that some vertex $y$ of $C$ is
black.  By Claim~\ref{cl2} (3), the predecessor $x\in V(C)$ of $y$ is
not white, i.e., it is in $B\setminus W$.  Applying Claim~\ref{cl2}
(3) repeatedly we obtain that every vertex of $C$ is in $B\setminus
W$.  So $C$ is monochromatic.  Thus Claim~\ref{cdm} holds.


\begin{claim}
     \label{aunic}
If a component $A$ of $R$ contains a cycle, then $A$ is unicyclic and
$|\crit(A)|=  |V(A)|=  |A|$.
\end{claim}
Let $A$ contain a cycle $C$.  By Claim~\ref{cdm} and by symmetry, $C$
is a black directed cycle.


For every edge $e= xy\in C$, by Claim~\ref{P3}, we have $x\in
\crit(e)$, and by Claim~\ref{noncritical}, we have $d_{R}^{+}(x)= 1$.
Let $A_{0}$ be a subgraph of $A$ that is maximal with the following
property: $A_0$ contains $C$, for every $x\in V(A_{0})$ there is a
directed path in $A_{0}$ from $x$ to some vertex of $C$, and every
vertex $x\in V(A_{0})$ is black and satisfies $d_{R}^+(x)= 1$.  Let us
prove that $A_{0}= A$.  Note that $A_0$ exists, because $C$ itself
satisfies all the required properties.

Suppose that $A_0\neq A$.  Then, since $A$ is connected, there is a
vertex $x\in V(A)\setminus V(A_{0})$ that is adjacent to some $y\in
V(A_{0})$.  Because $d_{R}^+(y)= 1$ and $d_{A_{0}}^+(y)= 1$, we have
$e= xy\in A$.  Since $y$ is black, Claim~\ref{P3} implies $\crit(e)=
\{x\}$.  Observe that $A_{0}$ contains a directed $P_{3}$ from $y$
containing black vertices.  Hence by Claim~\ref{cl2} (3), we have
$y\in B\setminus W$.  If $x\in W\setminus B$, then let $f= ux\in R$ be
any white head arrow.  Define $R^{*}= (R\setminus \{e\}) \cup \{f\}$.
The digraph $D^{*}= D\setminus R^{*}$ belongs to $D(k-1, k-1)$,
because $d_{D^*}^+(y)= d_{D'}^+(y)\leq k-1$, and $d_{D^*}^-(x)\leq
k-1$.  This contradicts the choice of $R$.  So $x$ is black.  Because
$x$ is black and $x\in\crit(e)$, Claim~\ref{noncritical} implies
$d_{R}^+(x)= 1$.  Hence one could include $x$ to $A_{0}$,
contradicting the maximality of $A_0$.  Therefore $A_{0}= A$.

Since $A= A_0$, $C$ is the only cycle in $A$ and every vertex $x$ of
$A$ is black and satisfies $d_{R}^+(x)=  1$.  %
Claim~\ref{P3} implies that every vertex of $A$ is a critical vertex
of its outgoing edge.  %
So $|\crit(A)|=  |V(A)|=  |A|$, and Claim~\ref{aunic} holds.

\

Claims~\ref{atree} and~\ref{aunic} show that $|\crit(A)|\geq  |A|$ is
true for every connected component $A$.  If $A_1, \ldots, A_t$ are the
components of $R$, we have clearly $\crit(R)= \crit(A_1)\cup\cdots\cup
\crit(A_t)$.  Thus we obtain $|\crit(R)|\geq  |R|$, which proves
Claim~\ref{critr}.  This concludes the proof of the theorem.
\end{proof}

The regular tournament on $2k+1$ vertices has indegree equal to
outdegree for every vertex, hence it is in $D(k, k)$.  To obtain a
subgraph belonging to $D(k-1, k-1)$ one has to remove at least $k+1$
from its $m=  {2k+1\choose 2}$ edges.  This shows that the tournament
has no subgraph in $D(k-1, k-1)$ containing more than ${2k+1\choose
2}-k+1=  m-(1+1/k)m/(2k+1)=  (2k-1/k)m/(2k+1)$ edges.

\begin{corollary}
     \label{3/5}
Every digraph $D\in D(2, 2)$ with $m$ edges contains a subgraph
belonging to $D(1, 1)$ with at least $3m/5$ edges.\hfill $\Box$
   \end{corollary}

We note that for $k=  1$, Theorem~\ref{k-1} yields another proof that
every digraph $D\in D(1, 1)$ with $m$ edges contains a directed cut of
size at least $m/3$ (cf.  Corollary~\ref{m/3}).

\section{Cuts in $D(k, k)$}
\label{secd22}

In \cite{alonetal} it was observed that the $k$-regular orientation of
the complete graph on ${2k+1}$ vertices has no directed cut of size
more than $(\frac{1}{4}+\frac{1}{8k+4}){2k+1\choose 2}$.
Consequently, in a digraph $D\in D(k, k)$ with $m$ edges one cannot
guarantee a directed cut of size larger than $(\frac{1}{4}+
\frac{1}{8k+4})m$.  It was proved in \cite{alonetal} that every
digraph with outdegree at most $k$ does contain a directed cut of that
size.  Using the same methods we show that it is also true for the
acyclic members of $D(k, k)$.

The basic tool is a lemma in \cite{lehel} that is proved there by
elementary counting.

\begin{lemma}[\cite{lehel}]
     \label{counting}
If a $\gamma$-colorable graph $G$ has $m$ edges, then it has a
bipartite partial graph with at least
$(\lfloor\gamma^{2}/4\rfloor/{\gamma\choose 2})m$ edges.\hfill$\Box$
   \end{lemma}

\begin{theorem}
\label{acyclic}
If $D\in D(k, k)$ is acyclic and has $m$ edges, then $D$ contains a
directed cut of size at least $(\frac{1}{4}+\frac{1}{8k+4})m$.
\end{theorem}
\begin{proof}
Let $D^{+}$ be the subgraph of $D$ induced by the set $X= \{v\in V(D)|
d^{+}(v)\leq k\}$ and let $D^{-}$ be the subgraph of $D$ induced by
$V(D)\setminus X$.  Because $D$ is acyclic, every subgraph of $D^{+}$
has a source, thus its underlying graph $G^{+}$ is $k$-degenerate.
Similarly, every subgraph of $D^{-}$ has a sink, thus its underlying
graph $G^{-}$ is $k$-degenerate.  Consequently, both graphs $G^{+}$
and $G^{-}$ are $(k+1)$-colorable, therefore the underlying graph $G$
of $D$ is $(2k+2)$-colorable.

Applying Lemma~\ref{counting} with $\gamma= 2k+2$, we obtain a
bipartite partial graph of $G$ with $\frac{k+1}{2k+1}m$ edges.  In $D$
at least half of the edges of that bipartite graph form a directed cut
of size at least $\frac{k+1}{4k+2}m= (\frac{1}{4}+\frac{1}{8k+4})m$.
\end{proof}

We do not know whether Theorem~\ref{acyclic} remains true for all
digraphs in $D(k, k)$, and for every $k$.  The coefficients are $1/3$,
$3/10$ and $2/7$ for $k= 1, 2$, and $3$, respectively.  By Theorem
\ref{three}, a digraph $D\in D(1, 1)$ of size $m$ has a cut with at
least $m/3$ edges.  The theorem below answers the question
affirmatively for $D(2, 2)$.

\begin{theorem}
     \label{3/10}
Every digraph $D\in D(2, 2)$ with $m$ edges has a directed cut of size
at least $3m/10$.
\end{theorem}
\begin{proof}
We prove the theorem by induction on $m$.  For $m= 1$ the theorem is
true.  Now suppose that $m\ge 2$ and that the theorem holds for every
digraph with at most $m-1$ edges.  Since $D$ is in $D(2, 2)$, its
vertex set can be partitioned into two sets $X, Y$ such that every
vertex $x\in X$ satisfies $d^-(x)\le 2$ and every vertex $y\in Y$
satisfies $d^+(y)\le 2$.  Consider the set of edges $F = \{xy\in E|
x\in X, y\in Y\}$.

First suppose that the underlying bipartite graph $(X, Y; F)$ contains
a cycle.  Let $C$ be any such cycle, say with length $2k$, let $X_C$
and $Y_C$ be the set of vertices of $C$ that lie in $X$ and $Y$
respectively, and let $F_C$ be the set of edges of $C$.  So $|F_C| =
2k$.  Let $E_C$ be the set of edges such that either their end is in
$X_C$ or their origin is in $Y_C$.  By the definition of $X, Y$ and
the fact that $D$ is in $D(2, 2)$, we have $|E_C|\le 4k$.  Consider
the digraph $D' = D \setminus (E_C\cup F_C)$.  Clearly, $D'\in D(2,
2)$, and the number $m'$ of edges of $D'$ satisfies $m' \ge m -6k$.
By the induction hypothesis, $D'$ has a directed cut of size at least
$3m'/10$.  If the edges of $F_C$ are added to such a directed cut, we
obtain a directed cut of $D$, because $D'$ does not contain any edge
of $E_C$.  This directed cut of $D$ has size at least $3m'/10+|F_C|\ge
3(m-6k)/10 + 2k \ge 3m/10$.  So the theorem holds for $D$.

Now suppose that the bipartite graph $(X, Y; F)$ does not contain any
cycle.  Thus $ |F| \le n-1$, where $n$ is the number of vertices of
$D$.  By the definition of $X$ and $Y$, we have $m\le 2 |X| + 2|Y| +
|F| \le 2 n + n-1= 3n-1$, which implies that $D$ has a vertex $v$ of
degree at most $5$.  Actually the same argument can be repeated with
$D\setminus \{v\}$, and so on.  Thus the underlying graph of $D$ is
$5$-degenerate and therefore has chromatic number at most $6$.
Applying Lemma~\ref{counting} with $\gamma= 6$ we obtain a bipartite
subgraph with $3/5$ edges, thus $D$ has a directed cut of size at
least $3m/10$.
\end{proof}

\section{Problems}
\label{problems}

Let $c_{max}$ be the ratio of the maximum directed cut size
 to the edge count $m$ of a digraph.  For connected digraphs
of $D(1, 1)$, Theorem~\ref{7/20} improves the basic estimation
$c_{max}\geq 1/3$ to $c_{max}\geq 7/20$ provided $m>3$.  On the other
hand, Example 1 before Theorem~\ref{cutformula} shows infinitely many
connected digraphs of $D(1, 1)$ with $c_{max} < 3/8$.  We conjecture
that the bound $c_{max}\geq 7/20$ can be improved to $3/8$ in the
limit in the following sense.
\begin{problem}
For every $\varepsilon > 0$, there exists a constant $m_{\varepsilon}$
such that $c_{max} > 3/8-\varepsilon$ holds for every connected
digraph of $D(1, 1)$ with $m>m_{\varepsilon}$ edges.
\end{problem}

At some point of the investigation in $D(1, 1)$ we observed that the
presence of source or sink vertices of the digraph increases the size
of a maximum directed cut.  Corollary~\ref{notriangle} might have the
following sharpening.
\begin{problem}
If a connected digraph $D\in D(1, 1)$ with $m$ edges contains no
directed triangle and has $s$ vertices with indegree or outdegree
zero, then $D$ has a directed cut of size at least $(2m+s)/{5}$.
\end{problem}

Bondy and Locke \cite{bondyetal} proved that a triangle-free subcubic
graph has a cut (a bipartite subgraph) of size at least $4m/5$.  A
characterization of all extremal graphs for that bound was given by Xu
and Yu in \cite{xuetal}.  The problem of characterizing the extremal
graphs for the bound of Corollary~\ref{notriangle} remains open:
\begin{problem}
Determine the list of all digraphs $D\in D(1, 1)$ of size $m$ that
contain no directed triangle and have no directed cut with more than
$2m/5$ edges.
\end{problem}

Bondy and Locke's result in \cite{bondyetal} consists of a polynomial
time algorithm that finds a cut with at least $4m/5$ edges in a
triangle-free subcubic graph.  It is known that finding a maximum cut
is NP-hard even in the restricted family of triangle-free cubic graphs
(see Yannakakis \cite{yan}).  Even the approximation of the max cut
problem in cubic graphs within the ratio of $0.997$ is NP-hard (see
Berman and Karpinski \cite{bekar}).  On the other hand, Halperin,
Livnat and Zwick \cite{HLZ} give a polynomial time approximation
algorithm with ratio $0.9326$.

Concerning digraphs in $D(1,1)$, Corollary~\ref{alg720} gives a
polynomial time algorithm that produces a cut of size at least $7m/20$
in every digraph in $D(1,1)$ of which no component is a directed
triangle; and so this is an approximation algorithm with ratio $0.35$.
Can a better ratio be obtained?  Actually, as far as we know, none of
the known results implies that computing the exact value of a maximum
directed cut is NP-hard in $D(1,1)$.  So we ask:
\begin{problem} 
What is the complexity status of computing the size of a maximum
directed cut in a digraph of $D(1,1)$?  If it is NP-hard, what is the
best value of $\varepsilon$ for which there is a polynomial time
approximation algorithm with ratio $1-\varepsilon$ for this problem?
\end{problem} 
The same problem can be posed for digraphs of $D(1,1)$ with no 
directed triangle, or with no triangle at all. 

\

How large a subgraph belonging to $D(1, 1)$ can be found in a digraph
$D\in D(2, 2)$?  Corollary~\ref{3/5} says that $D$ with $m$ edges
contains a subgraph in $D(1, 1)$ with at least $3m/5$ edges.  This
lower bound is probably not sharp.
\begin{problem}
Determine the largest constant $\lambda$ such that in every digraph
$D\in D(2, 2)$ with $m$ edges there exists a subgraph $D'\in D(1, 1)$
of size at least $\lambda m$.
\end{problem}
If $D$ is the regular tournament on five vertices, then $D\in D(2,2)$
and one needs to remove at least three edges to obtain a subgraph
$D'\in D(1, 1)$.  This shows that in the problem above $\lambda\leq
7/10$.

\

From a result in \cite{alonetal} it follows that the edges of every
graph $D\in D(2, 2)$ can be decomposed into at most five directed
cuts.  Furthermore, four cuts are sufficient if $D$ is acyclic.  The
regular tournament on five vertices shows that four cuts might be
necessary.  Indeed, it has $10$ edges, and the size of a directed cut
is at most $3$.  No example has been found to show that five directed
cuts are necessary.

\begin{problem}
The edges of every digraph $D\in D(2, 2)$ can be decomposed into at
most four directed cuts.
\end{problem}

Several problems remain open in $D(2, 2)$ pertaining to the ratio
$c_{max}$.

\begin{problem}
If $D\in D(2, 2)$ has $m$ edges and contains no copy of the regular
tournament on five vertices, then $D$ has a directed cut of size at
least $m/3$.
\end{problem}

We do not know whether Theorem~\ref{acyclic} pertaining to acyclic
digraphs remains true for all digraphs in $D(k, k)$, and for every
$k$.  The coefficients $\frac{1}{4}+\frac{1}{8k+4}$ are equal to
$1/3$, $3/10$, and $2/7$ for $k= 1, 2$, and $3$, respectively.  By
Corollary \ref{m/3}, and by Theorem~\ref{3/10}, a digraph $D$ of size
$m$ has a cut with $m/3$ and $3m/10$ edges, respectively for $D\in
D(1, 1)$ and $D\in D(2, 2)$.  The next case $k= 3$ is proposed here as
a question.  It is quite possible that the answer is negative.  Even
if it is not the case we conjecture that Theorem~\ref{acyclic} does
not extend for every $k$.

\begin{problem}
Is it true that every digraph of $D(3, 3)$ with $m$ edges contains a
directed cut of size at least $2m/7$?
\end{problem}

Digraphs with maximum outdegree $k$ satisfy $c_{max}\geq \frac{1}{4}+
\frac{1}{8k+4}$, and this is the best bound, as shown in
\cite{alonetal}.  It is worth noting that the same bound was obtained
here in Theorem~\ref{acyclic} for acyclic members of $D(k, k)$.
Furthermore, the regular tournament on $2k+1$ vertices is an example
of a digraph with no directed cut larger than $(\frac{1}{4}+
\frac{1}{8k+4})m$.  We believe that in the larger family $D(k, k)$
there are more examples showing that this bound cannot be achieved,
provided $k$ is large enough.

\begin{problem}
There exists a $k_{0}$ such that for every $k\geq k_{0}$ there are
digraphs in $D(k, k)$ with $c_{max} <\frac{1}{4}+\frac{1}{8k+4}$.
\end{problem}

In Theorem~\ref{k-1} we are dealing with the largest subgraph of $D\in
D(k, k)$ that belongs to the ``lower'' class $D(k-1, k-1)$.  This
leads naturally to the investigation of the minimum sets $R\subset
E(D)$ to be removed from $D$ in order to lower its class.  The proof
of Theorem~\ref{k-1} suggests that such minimum sets considered as
digraphs have a particular structure reminiscent of forests.
Repeating the procedure, one obtains a decomposition of the original
digraph $D\in D(k, k)$ into at most $k$ of these structures.

Practical applications motivate the study of decompositions of
digraphs into directed stars (see \cite{brandtetal}).  The
\emph{directed star arboricity} (dst) introduced in \cite{guiduli} is
defined as the minimum number of outstar forests (also called galaxy)
the edge set of a digraph can be partitioned.  For instance it is
proved in \cite{aminietal} that a digraph $D$ with indegree at most
$k$ has a decomposition into $k$ outforests plus one galaxy.  This
result implies $dst(D)\leq 2k+1$, and it is conjectured in
\cite{aminietal} that $2k$ is the tight bound, for $k\geq 2$.

As a general problem we propose here a similar decomposition theory of
the digraphs of $D(k, k)$ into appropriate forest-like structures.

\section*{Acknowledgment}

The first author's participation in the present research was supported
by the Professional Development Assignment obtained from The
University of Memphis and by the funding of the Universit\'e de
Grenoble 1, Joseph Fourier.  He is grateful for the help and the
hospitality of all colleagues at the Laboratoire Leibniz - IMAG during
his stay in Grenoble in October and November 2006.

This research has been supported by the ADONET network of the European
Community, which is a Marie Curie Training Network.

\end{document}